\documentclass{article}

\usepackage{PRIMEarxiv}
\usepackage{amsmath}
\usepackage[utf8]{inputenc} 
\usepackage[T1]{fontenc}    
\usepackage{hyperref}       
\usepackage{url}            
\usepackage{booktabs}       
\usepackage{amsfonts}       
\usepackage{nicefrac}       
\usepackage{microtype}      
\usepackage{lipsum}
\usepackage{fancyhdr}       
\usepackage{graphicx}       
\usepackage{bbm}
\usepackage{makecell}
\graphicspath{{media/}}     

\usepackage{algorithm}
\usepackage{algpseudocode}

\usepackage{epstopdf}
\AppendGraphicsExtensions{.tif}

\usepackage{mathtools}
\DeclarePairedDelimiterX{\infdivx}[2]{(}{)}{%
  #1\;\delimsize\|\;#2%
}

\title{Portfolio optimization using predictive auxiliary classifier generative adversarial networks with measuring uncertainty}

\author{
  Jiwook Kim \\
  School of Electrical \& Electronics Engineering \\
  Chung-Ang University \\
  Seoul, Korea\\
  \texttt{tom919@cau.ac.kr} \\
   \And
  Minhyeok Lee \\
  School of Electrical \& Electronics Engineering \\
  Chung-Ang University \\
  Seoul, Korea\\
  \texttt{mlee@cau.ac.kr} \\
}

\begin{document}
\maketitle

\begin{abstract}
In financial engineering, portfolio optimization has been of consistent interest. Portfolio optimization is a process of modulating asset distributions to maximize expected returns and minimize risks. To obtain the expected returns, deep learning models have been explored in recent years. However, due to the deterministic nature of the models, it is difficult to consider the risk of portfolios because conventional deep learning models do not know how reliable their predictions can be. To address this limitation, this paper proposes a probabilistic model, namely predictive auxiliary classifier generative adversarial networks (PredACGAN). The proposed PredACGAN utilizes the characteristic of the ACGAN framework in which the output of the generator forms a distribution. While ACGAN has not been employed for predictive models and is generally utilized for image sample generation, this paper proposes a method to use the ACGAN structure for a probabilistic and predictive model. Additionally, an algorithm to use the risk measurement obtained by PredACGAN is proposed. In the algorithm, the assets that are predicted to be at high risk are eliminated from the investment universe at the rebalancing moment. Therefore, PredACGAN considers both return and risk to optimize portfolios. The proposed algorithm and PredACGAN have been evaluated with daily close price data of S\&P 500 from 1990 to 2020. Experimental scenarios are assumed to rebalance the portfolios monthly according to predictions and risk measures with PredACGAN. As a result, a portfolio using PredACGAN exhibits 9.123\% yearly returns and a Sharpe ratio of 1.054, while a portfolio without considering risk measures shows  1.024\% yearly returns and a Sharpe ratio of 0.236 in the same scenario. Also, the maximum drawdown of the proposed portfolio is lower than the portfolio without PredACGAN, which demonstrates the effectiveness of the proposed methods. The proposed framework with PredACGAN and the proposed algorithm are anticipated to contribute to the expansion of discussions regarding deep learning-based portfolio optimization.

\end{abstract}


\section{Introduction}
Portfolio optimization is a method to find appropriate asset distributions that maximize portfolio returns and minimize the risks of the investment \cite{port1}. Since the construction of a robust portfolio has always been of great interest in financial engineering, portfolio optimization has been extensively studied in recent decades \cite{port_review, port2}. For portfolio optimization, many studies have employed historical returns and volatilities of an asset, assuming that expected future returns and volatilities of the asset will be similar to the historical data \cite{port_variance}. However, such an assumption is open to question and does not always accord with real-world data.

Therefore, the prediction methods for future returns have been widely studied as an alternative way to obtain the expected returns of an asset \cite{port_predict}. The underlying idea of these studies is that the predicted future returns provide a better estimation of expected returns than historical returns. Mainly, in recent years, deep learning models have been widely used to predict future asset returns, as the models have demonstrated superior performances in numerous problems and applications \cite{port_deep1, port_deep2, port_deep3, port_deep4, port_deep5, port_deep6, port_deep7}. For this reason, deep learning prediction-based portfolio optimization has attracted much attention.

However, such methods have an obvious limitation in that the model does not consider the risk of the portfolio. The deep learning models generally predict the expected returns, neglecting the uncertainty of the prediction. Since maximizing returns and minimizing risks are the two critical factors of portfolio optimization, it can be considered that these methods handle only half of the problem \cite{port_risk}. This limitation in the deep learning prediction-based portfolio optimization is mainly caused by the deterministic nature of deep learning models \cite{port_det}. A deep learning model consists of matrix multiplications and nonlinear functions; thus, the model conducts a point estimation for a given input. Since the point estimation does not tell how much the estimation is reliable, the risk of the portfolio based on the estimation can hardly be measured.

Conversely, probabilistic models provide the prediction as a form of distribution. Therefore, the variance of the predicted distribution can signify the uncertainty of the prediction, while the average or median can still function as the point estimation \cite{uncert_review}. Hence, probabilistic models can pursue both objectives of portfolio optimization, i.e., maximizing returns and minimizing risks, by considering the estimation variance as a risk. For example, if the variances of the predictions can be measured using the models, a simple strategy can be considered in which predictions with high variances are ignored and regarded as invalid; then, the portfolio is constructed only with valid predictions with low variances.

To address this problem, generative deep learning models have been investigated. While ordinary deep learning models are deterministic models, a generative deep learning model aims to provide probabilistic distributions as its output, enabling probabilistic prediction. Among the generative deep learning models, generative adversarial networks (GANs) \cite{gan} have been extensively studied \cite{gan_review}. Whereas the original GANs were introduced for sample generation, not the probabilistic prediction, Lee and Seok (2021) \cite{uncert_measure} have proposed a method to use a GAN for probabilistic prediction. They employed a conditional GAN \cite{cgan, cgan_proj} in which a sample is used for a condition, and a predictive distribution is produced, which can be interpreted as reversed inputs and outputs compared to the original auxiliary classifier GAN (ACGAN) \cite{acgan}. Then, the median and variance of the predictive distribution can be a prediction and a risk, respectively.

However, owing to the development of recent GANs, it has been studied that the conditional GAN used in Lee and Seok (2021) \cite{port_predict} does not demonstrate competitive performance compared to recent GANs \cite{acgan, spectral, wgan, hingegan}. Instead, the structure of ACGAN has been used in many recent applications. In ACGAN, a classification layer is additionally used with the original GAN, then the conditional input variable learns from the classification layer. Nevertheless, ACGAN has been employed mainly for sample generation; a prediction model using ACGAN has not been actively studied thus far.

In this study, a predictive ACGAN (PredACGAN) is proposed for portfolio optimization. In the proposed PredACGAN, the samples are used as conditional inputs to generate predictive distributions; thus, the predictive distributions become the prediction of the corresponding samples. The training process is similar to the original ACGAN, where a discriminator with a classification layer aims to learn the difference between generated distributions and real distributions, while a conditional generator produces the generated distributions to deceive the discriminator. However, in PredACGAN, the generator predicts future returns with estimation risks, whereas the original ACGAN generates synthetic samples that mimic real samples. 

To evaluate PredACGAN, 30 years of daily market price data of Standard and Poor's 500 (S\&P 500) stocks are used. A scenario of constructing a portfolio by selecting a certain percentage of stocks among the investment universe is supposed. To construct a portfolio using PredACGAN, a straightforward algorithm is proposed to select stocks with estimated risks; in the proposed algorithm, predictions with high risks are ignored, then the portfolio is composed only of predictions with low risks.

\section{Related work}
\subsection{Problem description}
Portfolio optimization is a process of selecting the best portfolios among possible combinations of asset distributions \cite{port1}. Therefore, the objective of portfolio optimization is to maximize investment returns with considering the minimization of risks. Conventional portfolio optimization methods generally have achieved the objective by modulating the asset distributions among the investment universe \cite{port_review}. This process can be simplified as follows:

\begin{equation}\label{eq:po}
W = \underset{W = \{w_1,...,w_N\}}{\arg\max} \sum_{k}^{}{w_k E(r_k)} - \lambda_{r}R(W), 
\end{equation}
subject to 
\begin{equation}
\sum_{k}^{}{w_k = 1} \quad \textrm{and} \quad 0 \leq w_k \leq 1,
\end{equation}
where $w_k$ denotes the distribution weight of $k^{th}$ asset; $N$ indicates the number of assets in the investment universe; $E(r_k)$ is the expected return of $k^{th}$ asset; $\lambda_{r}$ is a weight parameter for the portfolio risk; $R(W)$ is a risk measurement of the portfolio with $W$. Therefore, appropriate estimations of $E(r_k)$ and $R(W)$ are one of the essential parts in portfolio optimization. Conventionally, an average historical return and historical volatility have been used for $E(r_k)$ and $R(W)$. For instance,  $E(r_k):=\frac{1}{T}\sum_{t=1}^{T}{r_{k,t}}$ and $R(W):=\sum_{k}^{}\sigma_{T}(w_{k} r_{k,t})$, where $T$ is the number of historical data; $r_{k,t}$ is a historical return of $k^{th}$ asset at time $t$; $\sigma_{T}(\cdot)$ is a standard deviation.

However, an alternative approach with prediction methods has been explored for the calculation of $E(r_k)$. The fundamental concept of such methods is to replace the historical data with predictions with statistical and machine learning models \cite{port_ml, port_ml2, port_ml3}. Among these methods, deep learning models have been extensively studied as they have shown outstanding performance in numerous other domains. In deep learning models, $E(r_k)$ is predicted by various data types as their inputs that have been considered to be related to future returns, such as technical indicators of asset prices and news articles. Such a concept can be represented as follows:

\begin{equation}
\hat{E}(r_{k,t}) := f_{\hat{\theta}}(X_{k,t-m}), 
\end{equation}
where $f_{\hat{\theta}}$ indicates a trained deep learning model; $X_{k,t-m}$ represents available input data for the model at time $t$ which is usually current time.

While deep learning models have demonstrated promising results, however, most studies have generally focused on the prediction of future returns and rarely considered $R(W)$, which is another key objective of portfolio optimization \cite{port_deep2, port_deep4}. In portfolio optimization as well as actual investment, risk management is considered the most crucial. Thus, conventional prediction-based methods have shown a limitation in that they can hardly consider the risk of portfolios. This limitation is due to the deterministic nature of conventional deep learning models where $f_{\hat{\theta}}$ can simply be considered as matrix multiplications with nonlinear functions. Therefore, for a given $X_{k,t-m}$, the prediction $\hat{E}(r_{k,t})$ is a scalar value which corresponds to a point estimation.

This study aims to propose a probabilistic deep learning model that targets to predict both $E(r_k)$ and $R(W)$ in Eq. \ref{eq:po} to address the limitation of the conventional deep learning-based model. In the proposed model, multiple numbers of $E(r_k)$ are predicted for given $X_{k,t-1}$, thereby making a form of distribution. Then, the risk of predictions is measured by the entropy of the predicted distribution. The framework of the proposed model can be represented as follows:

\begin{equation}\label{eq:distwithmodel}
\mathcal{D}_{k,t}=\{\hat{E}(r_{k,t,1}), ..., \hat{E}(r_{k,t,i}), ..., \hat{E}(r_{k,t,I})\} := f_{\hat{\theta}}(X_{k,t-m}),
\end{equation}
\begin{equation}\label{eq:estwithmodel}
\hat{E}(r_{k,t}):=Med(\mathcal{D}_{k,t}),
\end{equation}
\begin{equation}
\mathcal{R}_{k,t}:=Entropy(\mathcal{D}_{k,t}), \quad R(W):=\{\mathcal{R}_{1,t},...,\mathcal{R}_{N,t}\},
\end{equation}
where $\mathcal{D}_{k,t}$ is a set of predictions for the expected returns of $k^{th}$ asseat at time $t$; $I$ is the number of predictions conducted by the model with the same input of $X_{k,t-m}$; $\mathcal{R}_{k,t}$ indicates the measured risk; $Med(\cdot)$ denotes the median of a set.

\subsection{Auxiliary classifier generative adversarial networks}\label{sec:acgan}
GAN is a framework using neural network (NN) architectures, which has been introduced as a generative deep learning model. The conventional GAN is composed of two NN architectures, a generator and a discriminator. Two NNs play a two-player minimax game, where the discriminator aims to detect the output of the generator compared to real samples, whereas the generator targets to deceive the discriminator. Therefore, competitive learning is conducted between the generator and discriminator. As a result of the learning, the generator can produce synthetic samples which mimic features of the real samples by mapping the feature distribution onto the input of the generator. In this manner, GAN has been generally used for synthetic sample generation since the model can produce realistic samples \cite{gan, dcgan, infogan}. Such a training process can be interpreted as a minimax game with an objective function as follows:

\begin{equation}\label{eq:gan}
\min_{D}\max_{G} O(D,G):=\mathbb{E}_{x}\Bigl[L_{r}(D(x))\Bigr] + \mathbb{E}_{z}\Bigl[L_{f}(D(G(z)))\Bigr],
\end{equation}
where $O(D,G)$ is the objective function of the minimax game; $D$ and $G$ indicate the discriminator and generator, respectively; $x$ and $z$ are a sample and a random vector sampled from a dataset and  a distribution, respectively; $L_{r}$ and $L_{f}$ are the loss functions for real samples and fake samples, respectively.

Although produced synthetic samples by vanilla GANs show a high feature-wise similarity to real samples, there is a limitation in that the features of synthetic samples cannot be controlled. For instance, if a handwritten digit image dataset is trained with a GAN, the model may produce realistic handwritten digit images; however, the produced digits are random from zero to nine, and one cannot specify a specific digit to be produced. Accordingly, features in a produced sample cannot be controlled in vanilla GANs.

To address this limitation, the auxiliary classifier GAN (ACGAN) model has been proposed. In ACGAN, the discriminator has an auxiliary classification layer and classifies classes of samples (e.g., digits). The generator uses an encoding input (e.g., a one-hot encoding vector with ten elements) to represent the classes of samples, then learns from the classification layer in the discriminator. After training, one can control the produced samples by modifying the encoding input since feature information in the classes has been mapped onto the encoding input. The structure of the ACGAN model and its training process converts the objective function in Eq. \ref{eq:gan} to be as follows:

\begin{equation}\label{eq:acgan1}
\min_{D} O(D,\hat{G}) := \mathbb{E}_{x|c}\Bigl[L_{r}(D(x|c)) + \lambda_{cD}L_{c}(D_{c}(x|c),c)\Bigr] + \mathbb{E}_{z,c}\Bigr[L_{f}(D(\hat{G}(z,c)))\Bigr],
\end{equation}
\begin{equation}\label{eq:acgan2}
\max_{G} O(\hat{D},G) := \mathbb{E}_{z,c}\Bigl[L_{f}(\hat{D}(G(z,c))) - \lambda_{cG}L_{c}(\hat{D}_{c}(G(z,c)),c)\Bigr],
\end{equation}
where $c$ is the class information; $\lambda_{cD}$ and $\lambda_{cG}$ are hyperparameters to modulate the class leaning for the discriminator and generator, respectively; $L_{c}(\cdot,\cdot)$ is a loss function for classification; $D_{c}$ is the classification layer output of the discriminator.

\section{Methods}
\label{sec:headings}

This study proposes PredACGAN, a framework using ACGAN as a predictive model, whereas the original ACGAN is not be used as a predictive model. The fundamental idea of the proposed model is to reverse the inputs and outputs of the original ACGAN. In the original ACGAN, sample data are generally used for the input of the discriminator; however, samples are employed as conditional input in PredACGAN. Let $X_{k,t-m}$ be available input data at time $t$ to predict future returns of $k^{th}$ asset. Also, since the future returns are given in a historical market price dataset, it is possible to compose a training dataset of ordered pairs, $(X_{k,t-m}, r_{k,t})$, where $X_{k,t-m}:=\left[ (P_{k,t-m-1} - P_{k,t-m}) / P_{k,t-m}, \cdots , (P_{k,t-m-T_{i}} - P_{k,t-m}) / P_{k,t-m} \right]$ and $r_{k,t}:=(P_{k,t+T_{o}} - P_{k,t}) / {P_{k,t}}$, where $P$ represents the price of an asset, $T_{i}$ is the number of elements in $X_{k,t-m}$, and $T_{o}$ is a time interval of interest that the model aims to predict. Then, PredACGAN is trained with the following objective functions, which are modifications of Eqs. \ref{eq:acgan1} and \ref{eq:acgan2}:

\begin{align}\label{eq:pacgan1}
\min_{D} O(D,\hat{G}) := & \mathbb{E}_{r_{k,t}}\Bigl[L_{r}(D(r_{k,t})) + \lambda_{cD}L_{c}(D_{c}(r_{k,t}),X_{k,t-m})\Bigr] \nonumber\\
& + \mathbb{E}_{z,X_{k,t-m}}\Bigl[L_{f}(D(\hat{G}(z,X_{k,t-m})))\Bigr],
\end{align}
\begin{equation}\label{eq:pacgan2}
\max_{G} O(\hat{D},G) := \mathbb{E}_{z,X_{k,t-m}}\Bigl[L_{f}(\hat{D}(G(z,X_{k,t-m}))) - \lambda_{cG}L_{c}(\hat{D}_{c}(G(z,X_{k,t-m})),X_{k,t-m})\Bigr].
\end{equation}

Such a concept of PredACGAN can be interpreted that $X_{k,t-m}$ and $r_{k,t}$ are used as a condition and a generated sample, respectively. Therefore, it can also be said that the model can generate "a prediction sample" of $r_{k,t}$ with "a class condition" of $X_{k,t-m}$, which eventually corresponds to prediction of $r_{k,t}$. This method in PredACGAN is distinct from the original ACGAN in which ACGAN may target to generate $X_{k,t-m}$, given ordered pairs of $(X_{k,t-m}, r_{k,t})$.

After training of $G$ and $D$ with Eqs. \ref{eq:pacgan1} and \ref{eq:pacgan2}, the generator can produce multiple predictions with the same input of $X_{k,t-m}$. Let $z_{i} \sim \mathbb{P}_{z}$ be a noise vector, which is one of the inputs of the generator. Then, a distribution can be composed as Eq. \ref{eq:distwithmodel}, with predictions with different noise vectors. Specifically, the generator takes $I$ number of different $z_{i}$ and produces predictions $I$ times with respect to $X_{k,t-m}$:

\begin{align}\label{eq:d}
    \mathcal{D}_{k,t}&=\{\hat{G}(z_{1},X_{k,t-m}), \hat{G}(z_{2},X_{k,t-m}), \cdots, \hat{G}(z_{I},X_{k,t-m})\} \\
    &=\{ \hat{r}_{k,t,1},\hat{r}_{k,t,2}, \cdots, \hat{r}_{k,t,I} \}.
\end{align}

Then, the expected return with $X_{k,t-m}$ is predicted using the median of $\mathcal{D}_{k,t}$, as shown in Eq. \ref{eq:estwithmodel}. Finally, the expected return, $\hat{E}(r_{k,t})$, can be used for portfolio optimization process, as in Eq. \ref{eq:po}.

While it is true that the target data in the training set, i.e., $r_{k,t}$ in $(X_{k,t-m}, r_{k,t})$, can be a continuous variable with $r_{k,t}:=(P_{k,t+T_{o}} - P_{k,t}) / {P_{k,t}}$, the classification layer in ACGAN structures cannot handle such a datatype. Hence, $r_{k,t}$ is discretized to synthesize a categorical variable $C_{k,t}$ with three possible categories as follows:

\begin{equation}\label{eq:c}
C_{k,t} := 
    \begin{cases}
      c_{-} & \text{if \quad $r_{k,t} < Th_{l}$}\\
      c_{0} & \text{if \quad $Th_{l} \leq r_{k,t} < Th_{u}$}\\
      c_{+} & \text{if \quad $Th_{u} \leq r_{k,t}$}
    \end{cases},    
\end{equation}
where $Th_{l}$ and $Th_{u}$ are thresholds to discretized the target data. Thus, the training set for ACGAN is composed of ordered pairs of $(X_{k,t-m}, C_{k,t})$. Then, in the training process, the categorical target data are converted into one-hot vectors and used to obtain the classification loss in the classification layer of discriminator. 

\subsection{A risk measure with multiple predictions of PredACGAN}
In portfolio optimization, risks can be represented in various manners. For instance, conventional portfolio optimization methods regard the historical volatilities of an asset as a risk of the asset. Thus, given the historical volatilities of assets, conventional portfolio optimization methods decrease the risk of a portfolio by reducing the portfolio weights of the assets with higher historical volatility. However, it is still controversial that the future volatility of an asset accords with its historical volatility and historical volatilities can properly represent the risk of a portfolio.

In this study, a risk measure based on the uncertainty of predictions is introduced. In the measure, the entropy of $\mathcal{D}_{k,t}$ predicted by PredACGAN is used to calculate the risk:

\begin{alignat}{2}
    U_{k,t} & := && -D_{KL}\infdivx*{\mathbb{P}(C_{k,t}=c_{m})}{1-\mathbb{P}(C_{k,t}=c_{m})} \nonumber\\
            &   && - D_{KL}\infdivx*{1-\mathbb{P}(C_{k,t}=c_{m})}{\mathbb{P}(C_{k,t}=c_{m})} \\
            & = && -\sum_{i} \mathbb{P}(C_{k,t,i}=c_{m}) \cdot \log\Bigr(\frac{\mathbb{P}(C_{k,t,i}=c_{m})}{1-\mathbb{P}(C_{k,t,i}=c_{m})}\Bigl)\nonumber\\\label{eq:kl}
            &   && -\sum_{i} (1-\mathbb{P}(C_{k,t,i}=c_{m})) \cdot \log\Bigr(\frac{1-\mathbb{P}(C_{k,t,i}=c_{m})}{\mathbb{P}(C_{k,t,i}=c_{m})}\Bigl)
\end{alignat}
where $D_{KL}$ indicates the KL divergence; $C_{k,t,i}$ is an estimation of the generator with $(Z_{i}, X_{k,t-m})$; $U_{k,t}$ is the risk measure of the prediction with $X_{k,t-m}$; $c_{m} \in \{ c_{-}, c_{0}, c_{+} \}$ denotes the category that has a maximum probability among the possible categories, which is obtained as follows:

\begin{align}
    c_{m}:= \underset{C_{k,t}}{\arg\max} \Bigl\{ &\mathbbm{1}_{\{c_{-}\}}(C_{k,t}) \cdot \sum_{i}\mathbb{P}(C_{k,t,i}=c_{-}) +\mathbbm{1}_{\{c_{0}\}}(C_{k,t}) \cdot \sum_{i}\mathbb{P}(C_{k,t,i}=c_{0})  \nonumber\\
    &+\mathbbm{1}_{\{c_{+}\}}(C_{k,t}) \cdot \sum_{i}\mathbb{P}(C_{k,t,i}=c_{+}) \Bigr\},
\end{align}
where $\mathbbm{1}$ is the indicator function. 

This risk measure is based on a hypothesis that the features with high frequencies can be easily learned by the conditional input of the generator of PredACGAN, whereas the features with low frequencies are trained by the noise input of the generator. Specifically, it can be interpreted that $G: (\mathbb{P}_{z}, \mathbb{R}^{T_{i}}) \rightarrow \mathbb{R}^{3}$; therefore, the backpropagation-based training process of PredACGAN reinforces the weight parameter values connected to $X_{k,t-m} \in \mathbb{R}^{T_{i}}$ only if there are a strong relationship between $X_{k,t-m}$ and $C_{k,t,i}$; otherwise, $C_{k,t,i}$ may be mapped onto $z_{i} \in \mathbb{P}_{z}$.

Due to these properties of NN, a highly probable $X_{k,t-m}$ cannot cause much differences in $\mathbb{P}(C_{k,t,i})$ with respect to different values of $z_{i}$, resulting in stability of predictions. This implies that value of $\mathbb{P}(C_{k,t,i}=c_{m})$ in Eq. \ref{eq:kl} stable with respect to $z_{i}$, and accordingly, the value of $U_{k,t}$ becomes low, which represent a low prediction uncertainty and risk. In a similar manner, $U_{k,t}$ becomes high for a difficult-to-predict $X_{k,t-m}$.

\subsection{Portfolio weighting algorithm with the risk measure}\label{sec:algo}
In this section, a direct algorithm is proposed to select portfolios with the predictions and risk measures using PredACGAN. To use the risk measure, one of the direct methods is to eliminate the $k^{th}$ asset from the investment universe if $U_{k,t}$ is high. Based on the concept, the algorithm first select a specific percentage of assets among the the investment universe by using $\hat{C}_{k,t} = \mathbb{E}\left[\hat{G}(\{z_{i}\}_{1}^{I},X_{k,t-m})\right]$. Then, if $U_{k,t}$ of the selected $k^{th}$ asset is above a certain threshold, the asset is eliminated from the selection. For the selected assets, portfolio weights are equally distributed. The detailed process is shown in Algorithm \ref{alg:po}.

\begin{algorithm}
\caption{Portfolio weighting algorithm with PredACGAN (market-neutral position)}\label{alg:po}
\begin{algorithmic}[1]
\Require \textbf{(1)} $C = \{ \hat{C}_{1,t}, \cdots,  \hat{C}_{N,t}\}$: A set of predictions of PredACGAN for $N$ assets at time $t$; \textbf{(2)} $R = \{ U_{1,t}, \cdots, U_{N,t}\}$: A set of risk measures for $N$ assets at time $t$; \textbf{(3)} $Th_{p}$: A percentage threshold for the number of asset selection ($0 < Th_{p} \leq 0.5$); \textbf{(4)} $Th_{r}$: A risk threshold for the elimination of assets from the selection. 
\State \textbf{Objective:} Assign portfolio weights in $W=\{w_{1,t}, \cdots, w_{N,t}\}$.
\State \textbf{Initialize} $w_{k,t} \gets 0$, $\forall k \in \{1, \cdots, N \}$ and $n_{1}$, $n_{2} \gets 0$
\State \textbf{Calculate} scores $s_{k,t} \gets \mathbb{P}(\hat{C}_{k,t}=c_{+}) - \mathbb{P}(\hat{C}_{k,t}=c_{-})$, $\forall k$
\State \textbf{Calculate} index set $I_{w}=\{ i_{1}, \cdots, i_{N} \}$, such that $|I_{w}| = N$ and $s_{i_{k},t} \leq s_{i_{k+1},t}$, $\forall i_{k}, i_{k+1} \in I_{w}$
\For {$k \gets 1$ to $\lfloor N \times Th_{p} \rfloor$}
    \If{$U_{i_{k},t} < Th_{r}$}
        \State $n_{1} \gets n_{1}+1$
    \EndIf
\EndFor
\For {$k \gets \lfloor N \times (1-Th_{p}) \rfloor$ to $N$}
    \If{$U_{i_{k},t} < Th_{r}$}
        \State $n_{2} \gets n_{2}+1$
    \EndIf
\EndFor

\For {$k \gets 1$ to $N$}
    \If{$k \leq \lfloor N \times Th_{p} \rfloor$ and $U_{i_{k},t} < Th_{r}$}
        \State $w_{i_{k},t} \gets \frac{-1}{n_{1}}$
    \EndIf
    \If{$k \geq \lfloor N \times (1-Th_{p}) \rfloor$ and $U_{i_{k},t} < Th_{r}$}
        \State $w_{i_{k},t} \gets \frac{1}{n_{2}}$
    \EndIf
\EndFor
\State $W \gets \{w_{i_{1},t}, \cdots, w_{i_{N},t}\}$
\end{algorithmic}
\end{algorithm}

Specifically, first, a score for an asset is calculated in which the difference between probabilities of $c_{+}$ and $c_{-}$ is obtained. Thus, the score implies how much the price of the asset is likely to increase. Then, this score is used as the fundamental value for the portfolio construction; for instance, one can invest in the assets with high scores. However, a high score cannot assure the prediction is reliable. Therefore, the estimated risk measure of $k^{th}$ asset, i.e., $U_{k,t}$, is utilized to incapacitate the prediction. Using the estimated risk measure, one may invest only in the assets with high scores on conditions of low risks.

Algorithm \ref{alg:po} is a straightforward method to construct a portfolio with market-neutral position using this concept; therefore, $\sum_{k}w_{k,t} = 0$. Since the market-neutral position is supposed, a random portfolio or a uniform portfolio is likely to obtain zero return regardless of the market condition, including training data source, period of the investment, and specific market to be tested, which enable us to properly evaluate the proposed PredACGAN. Specifically, it is expected to obtain a high return only if the prediction is valid; also, even more, superior performance is expected if such an elimination process in Algorithm \ref{alg:po} with the estimated risk measure is effective. Hence, to evaluate the framework of PredACGAN, it can be considered to compare zero portfolio return (uniform portfolio weights), the return only with predictions, and the return with both predictions and risks.

\subsection{Training details of PredACGAN and experimental scenarios}\label{sec:training}
A few recent techniques for stable training of GAN are used since it is widely known that GAN easily diverges during training. First, the two-time-scale update rule (TTUR) \cite{ttur} method is applied to PredACGAN. In TTUR, Adam optimizer \cite{adam} is used with hyperparameters of $\beta_{1}=0.0$ and $\beta_{2}=0.9$, and learning rate for $D$ and $G$ are set to $2\times10^{-5}$ and $1\times10^{-5}$, respectively. The value of $\lambda_{cG}$ in Eq. \ref{eq:pacgan2} is set to 32. The hinge loss \cite{hingegan} is used for $D$, and Wasserstein distance \cite{wgan} is used for $G$, which correspond to $L_{r}$ and $L_{f}$ in Eqs. \ref{eq:pacgan1} and \ref{eq:pacgan2}. The models are trained with a computer system with RTX 3090 GPU. The batch size of the mini-batch algorithm is set to 16. The model is trained with 128 epochs. 

\begin{figure}
  \centering
  \includegraphics[width=\textwidth]{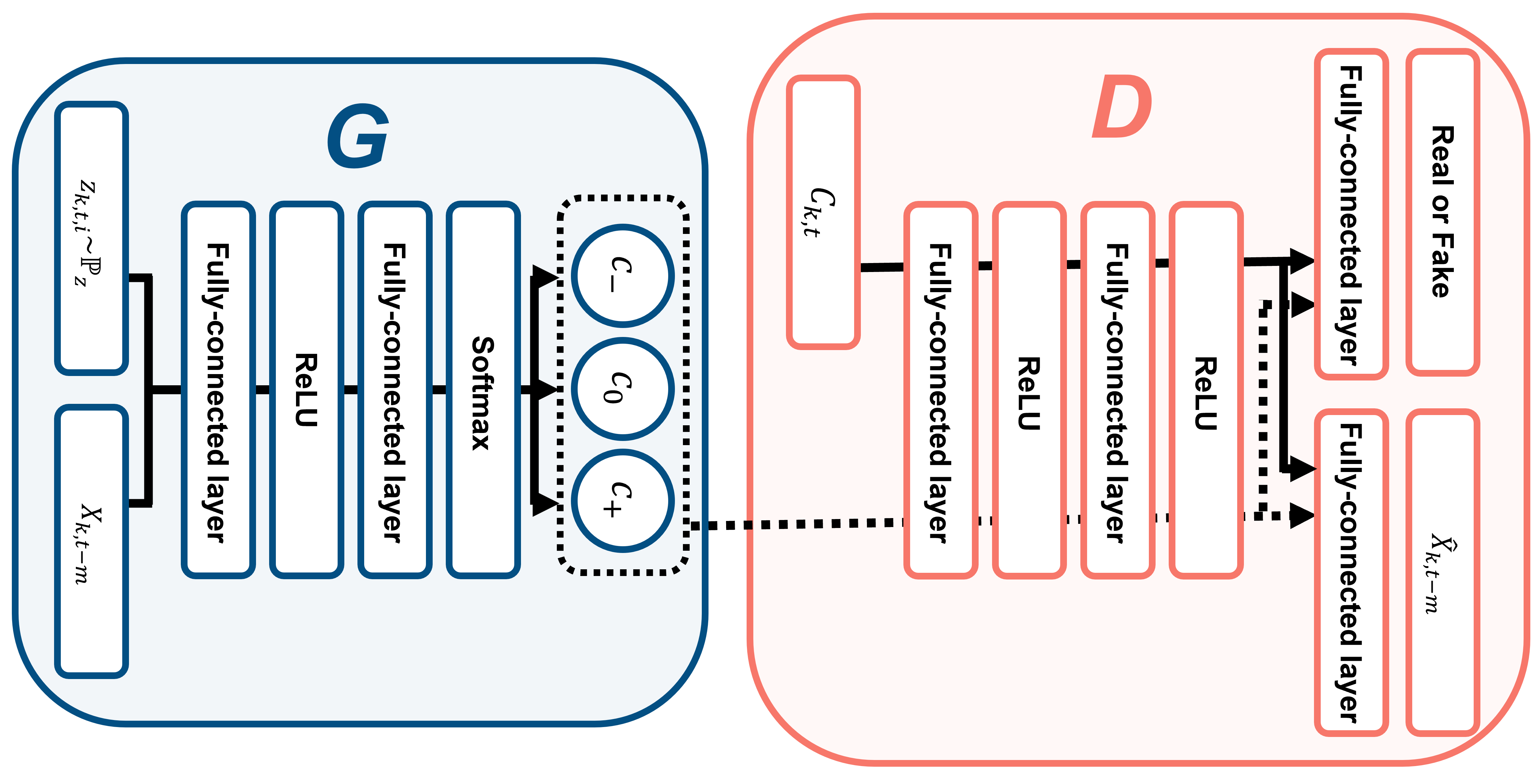}
  \caption{\textbf{The structure of PredACGAN.} All notations in this figure follow those of the paper.}
  \label{fig:fig1}
\end{figure}

For the NN structures, all layers in PredACGAN are fully-connected. $D$ is composed of two hidden layers, each of which has 2,048 nodes; $G$ consists of one hidden layer with 1,024 nodes. Parameter weight matrices are initialized by Xavier uniform initialization method. The rectified linear unit (ReLU) \cite{relu} is employed as the activation function of all hidden layers. The activation function of the output layer of $G$ is a softmax function because the probabilities of each category are estimated, and real samples are one-hot vectors. Since $G$ predicts a categorical distribution with three categories, as described in Eq. \ref{eq:c}, the number of output layer nodes of $G$ is set to three, each of which represents the categories of $C_{k,t}$. Also, because the classification in PredACGAN (i.e., $D_{c}$ in Eqs. \ref{eq:pacgan1} and \ref{eq:pacgan2}) aims to estimate $X_{k,t-m}$, the number of output layer nodes is $T_{i}$, which is the number of elements of $X_{k,t-m}$; for the same reason, $L_{c}$ in Eqs. \ref{eq:pacgan1} and \ref{eq:pacgan2} uses the $L_{2}$ loss. Since the hinge loss is used for $D$, the number of nodes of $D$ becomes one. For $\mathbb{P}_{z}$, $\prod_{128}N(0,1)$ is used. The number of sampling (i.e., $I$ in Eq. \ref{eq:d}) is set to 101, which is considered as a sufficient number to estimate the predictive distribution and calculate the median of the distribution. The structure of PredACGAN is illustrated in Figure \ref{fig:fig1}.

In the experiments of this study to evaluate PredACGAN, the experimental scenario is supposed to invest in S\&P 500 stocks; thus, the investment universe is 500 stocks in the index, and accordingly, $N=500$. Daily close market prices of the stocks from 1990 to 2020 are used for the experiment; the data from 1990 to 2010 are set to training data, and the data from 2011 to 2020 are set to evaluation data. The rebalancing period is set to one month, thereby $T_{o} \simeq 21$. In the scenario, returns from $t-1$ to $t-200$ of $k^{th}$ asset are used for $X_{k, t-m}$, where $t$ is a rebalancing moment; thus $T_{i} = 200$. The target data in the training set (i.e., $r_{k,t}$) are the return of $k^{th}$ assets from $t$ to $t+T_{o}$. Then, the target data are discretized, as described in Eq. \ref{eq:c}, with the threshold parameter values of $Th_{l}=-0.03$ and $Th_{u}=0.03$. The calculation of the training set, $X_{k, t-m}$ and $r_{k,t}$, can be represented as follows:

\begin{equation}
    X_{k,t-m}:=\left[ \frac{P_{k,t-1} - P_{k,t}}{P_{k,t}}, \cdots , \frac{P_{k,t-1-T_{i}} - P_{k,t}}{P_{k,t}} \right],
\end{equation}
\begin{equation}
    r_{k,t}:=\frac{P_{k,t+T_{o}} - P_{k,t}}{P_{k,t}},
\end{equation}
where $P_{k,t}$ is the daily close market price of $k^{th}$ asset at time $t$.

\section{Results \& Discussion}

\subsection{Experiments with stock market close prices in S\&P 500}
The proposed PredACGAN is evaluated with stock market close prices in S\&P 500 in this section \cite{S&P500_volatility}. A portfolio optimization scenario with a rebalancing period of one month is supposed. The detailed experimental scenarios are described in Section \ref{sec:training}. Additionally, as mentioned in Section \ref{sec:algo}, a performance comparison is performed between a portfolio with a uniform distribution of assets, a portfolio using only the predictions, and a portfolio with PredACGAN framework in which both the predictions and risk measure are employed with Algorithm \ref{alg:po}. Since the experiment is conducted with a market-neutral position with $\sum_{k}w_{k,t}=0$, the portfolio with a uniform distribution is expected to show zero return. 

Such an evaluation can properly assess PredACGAN framework, regardless of the experimental period and selection of specific markets for the experiment, since portfolio performances are highly dependent on such market conditions without the market-neutral position. To assess the portfolios, yearly and monthly return, Sharpe ratio \cite{sharpe}, information ratio \cite{IR}, and maximum drawdown are employed. In the portfolio construction, $Th_{p}$ is a hyperparameter to be chosen, which determines how many assets are invested. For instance, if $Th_{p}=10\%$ and $N=500$, the algorithm selects $500\times 10\% = 50$ assets with high scores for long positions and other $50$ assets with low scores for short positions. In the experiment, $Th_{p}=5\%$, $10\%$, and $20\%$ are tested.

\begin{table}
\centering
\caption{\textbf{Performance of PredACGAN.} $Th_{r}<0$ indicates portfolios without considering the risk measure. $Th_{p}$ is the threshold for the selection of assets. $Th_{r}$ is the threshold for the estimated risk measure. MMD: Maximum drawdown; IR: Information raio; SR: Sharpe ratio. Bold font represents the best performance; underline indicates the second-best. }\label{tab:comparision}

\begin{tabular}[width=\textwidth*0.95]{c|l|rrrrr}
\noalign{\smallskip}\noalign{\smallskip}\hline\hline
\multicolumn{1}{c|}{$Th_{p}$} & \multicolumn{1}{c|}{\textbf{Evaluation metrics}} &  \makecell{\multicolumn{1}{r}{$Th_{r}<0$} \\ \multicolumn{1}{r}{(Predictions only)}}& $Th_{r}<-10$ & $Th_{r}<-20$ & $Th_{r}<-30$ & $Th_{r}<-40$ \\
\hline

 & MMD & -0.122 & -0.125 & -0.281 & \underline{-0.121} & \textbf{-0.120}\\
 & IR & -0.274 & -0.066 & -0.008 & \textbf{0.319} & \underline{0.285} \\
5\% & Monthly SR & 0.057 & 0.059 & 0.034 & \textbf{0.331} & \underline{0.296} \\
 & Yearly SR & 0.236 & 0.207 & 0.327 & \underline{1.054} & \textbf{1.131} \\
 & Monthly Return (\%) & 0.081 & 0.082 &  0.113 & \textbf{0.867} & \underline{0.745}\\
 & Yearly Return (\%) & 1.024 & 0.094 & 3.217 & \textbf{9.123} & \underline{7.595}\\
\hline

 & MMD & \textbf{-0.085} & \underline{-0.105} & -0.282 & -0.121 & -0.120 \\
 & IR & -0.281 & -0.096 & 0.015 & \textbf{0.315} & \underline{0.285} \\
10\% & Monthly SR & 0.067 & 0.053 & 0.054 & \textbf{0.328} & \underline{0.296} \\
 & Yearly SR & 0.304 & 0.192 & 0.428 & \underline{1.058} & \textbf{1.131} \\
 & Monthly Return (\%) & 0.089 & 0.069 & 0.183 & \textbf{0.858} & \underline{0.745}\\
 & Yearly Return (\%) & 0.952 & 0.666 & 4.063 & \textbf{9.036} & \underline{7.595}\\
\hline

 & MMD & \textbf{-0.073} &\underline{-0.106} & -0.251 & -0.197 & -0.341 \\
 & IR & -0.275 & -0.066 & -0.028 & \textbf{0.175} & \underline{0.102} \\
20\% & Monthly SR & 0.097 & 0.062 & 0.017 & \textbf{0.195} & \underline{0.129} \\
 & Yearly SR & 0.340 & 0.221 & 0.227 & \textbf{0.646} & \underline{0.495} \\
 & Monthly Return (\%) & 0.129 & 0.087 & 0.058 & \textbf{0.606} & \underline{0.524}\\
 & Yearly Return (\%) & 1.185 & 0.825 & 2.177 & \textbf{5.540} & \underline{4.736}\\
\hline

\end{tabular}
\end{table}

\begin{figure}[htb!]
  \centering
  \includegraphics[width=\textwidth]{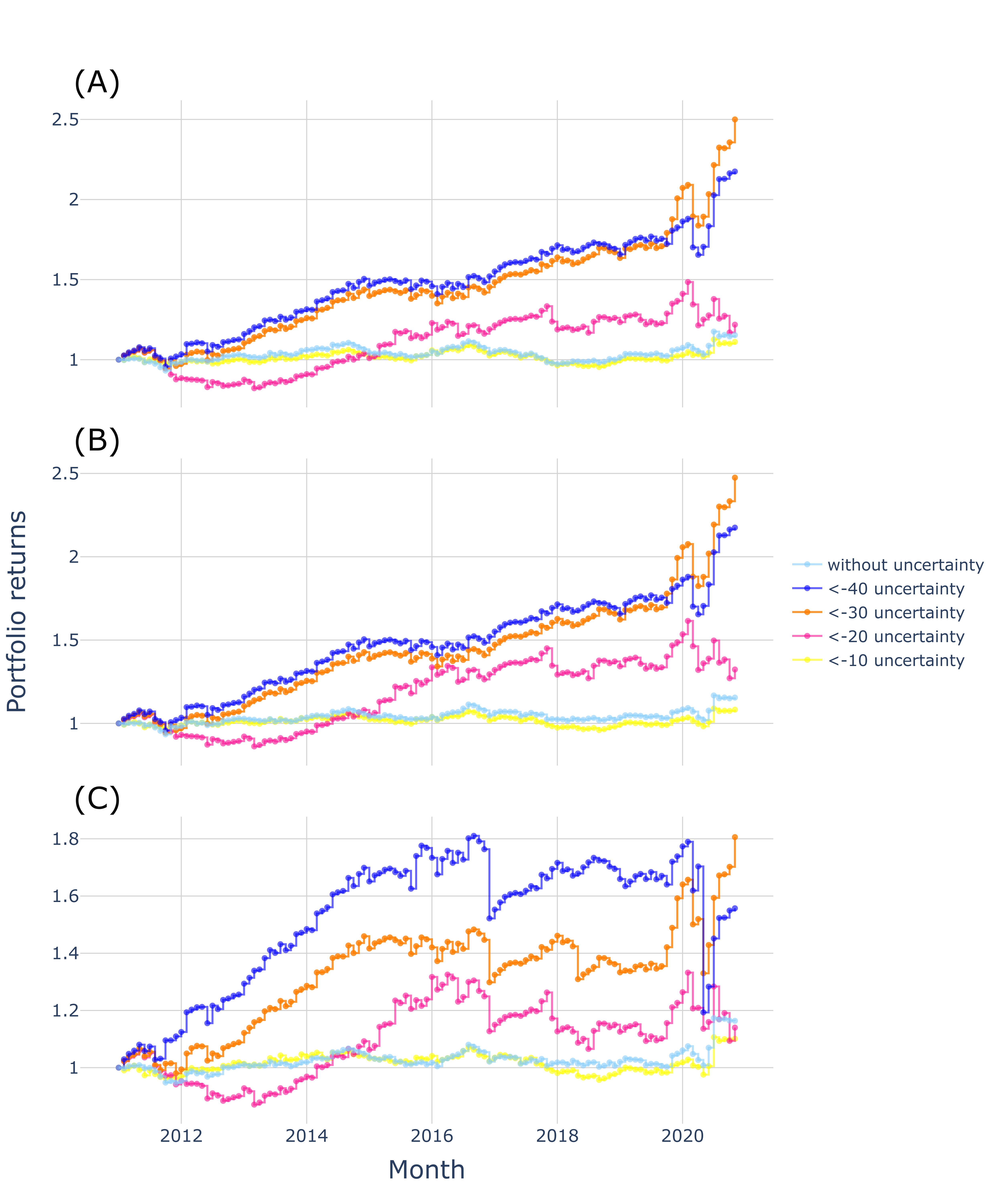}
  \caption{\textbf{Portfolio performance with respect to return during the evaluation period. (A)} Portfolios with $Th_{p}=5\%$. \textbf{(B)} Portfolios with $Th_{p}=10\%$. \textbf{(C)} Portfolios with $Th_{p}=20\%$. The 500 stocks in S\&P 500 index are the investment universe. The years from 2011 to 2020 are the evaluation period. Portfolios are composed with the market-neutral position, where a random portfolio is expected to obtain zero return.}
  \label{fig:pr}
\end{figure}

Table \ref{tab:comparision} exhibits the portfolio performance of PredACGAN with Algorithm \ref{alg:po}. For the return metrics, note that 0\% can be a baseline portfolio with uniform distribution of assets since the market-neutral position is assumed in this experiment. Also, $Th_{r}<0$ is the portfolio using only the predictions, which can be another baseline to evaluate the effectiveness of risk measures. As shown in the results, $Th_{r}<-30$ portfolios generally show the best performance, where they demonstrate the best results in 13 of the 18 measurements. The $Th_{r}<-40$ portfolios, which greatly consider the risks, demonstrate the second-best performance in general, in which they score the best results in three measurements and second-best in 14 measurements. These results indicate that the proposed framework using the estimated risk measure in PredACGAN is effective since considering risks shows superior performance compared to the uniform distribution of assets as well as the portfolio without the estimated risks.

Compared to the prediction-only portfolios under all conditions, the proposed portfolio demonstrates superior performances in terms of the Sharpe ratio that can measure returns by considering volatility. For instance, in the scenario with $Th_{p}=5\%$, the $Th_{r}<-30$ and $Th_{r}<-40$ portfolios show $347\%$ and $379\%$ improvements in terms of yearly Sharpe ratio. This result indicates the key advantage of the proposed methods since the Sharpe ratio can measure both return and risk. Also, in all scenarios, the $Th_{r}<-30$ portfolios exhibit the best performance in terms of monthly and yearly returns. For example, in the $Th_{p}=5\%$ scenario, the yearly return increases by $791\%$ compared to the baseline when the proposed methods are used.

Figure \ref{fig:pr} shows the performance with respect to return during the evaluation periods. Note that $1.0$ (100\%) return is the baseline since a market-neutral position is assumed. As shown in the results, the portfolios with $Th_{r}<-30$ and $Th_{r}<-40$ demonstrate the best and stable performance in $Th_{p}=5\%$ and $Th_{p}=10\%$, which indicates that considering the risk measure and eliminating high-risk assets with Algorithm \ref{alg:po} are valid. However, in $Th_{p}=20\%$, $Th_{r}<-30$ and $Th_{r}<-40$ portfolios show inferior performance compared to those in $Th_{p}=5\%$ and $Th_{p}=10\%$. As shown in Figure \ref{fig:pr}, this result is caused by an extraordinary case in early 2020, which corresponds to obvious outliers; still, the portfolios show better results compared to $Th_{r}<0$, which does not use the risk measure.

\subsection{Stability of training PredACGAN}

In this section, we discuss the stability of PredACGAN training. It is generally difficult to confirm whether GANs have trained properly since there is no clear method to validate it \cite{gan_stability}. In conventional NN structures, a loss gradually decreases, and such a convergence signifies fine training; however, in GANs, losses do not converge due to the nature of GAN training where competitive learning is conducted; if a module obtains a low value of the loss, the other module obviously has a high value of the loss. Conventionally, in GANs for image generation, there are several methods to validate the fine GAN training, such as Inception score \cite{Inception_score} and Frechet Inception distance (FID) \cite{ttur}. Since these methods can be used only for image data, PredACGAN cannot be evaluated these conventional methods.

\begin{figure}[htb!]
  \centering
  \includegraphics[width=\textwidth]{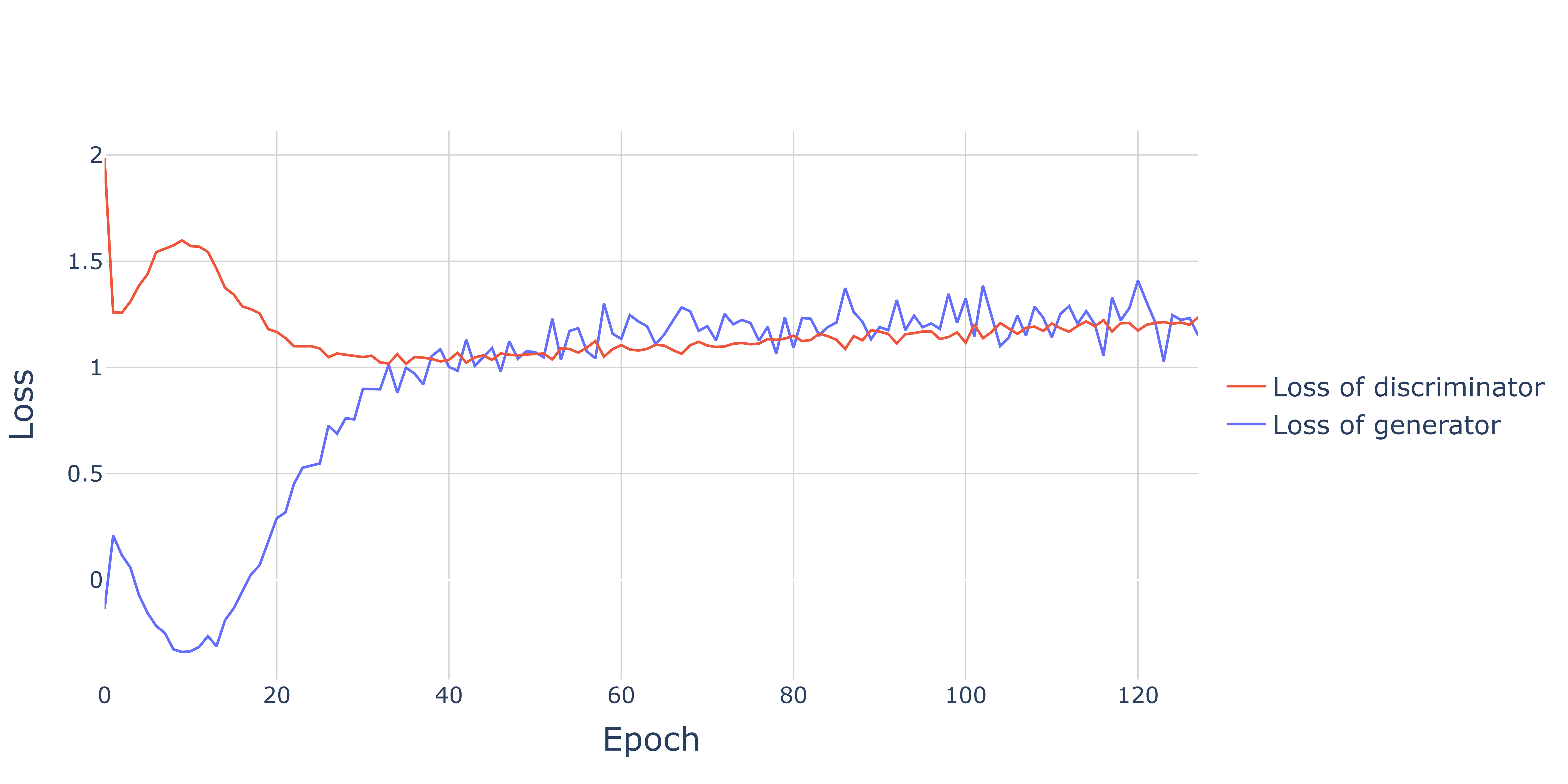}
  \caption{\textbf{PredACGAN training losses for the discriminator and generator with respect to epoch.}}
  \label{fig:loss}
\end{figure}

However, the superior performance shown in the experiment in this study reasonably explains that the training of PredACGAN was well conducted. Also, the loss values with respect to the epoch show that the training did not diverge. Although there is no precise method for loss values to evaluate whether the training is effective, at least the values tell whether the training diverged or not. As shown in Figure \ref{fig:loss}, the losses of $D$ and $G$ oscillate at approximately $1.0$; thus, it can be confirmed that the training was conducted without divergence. 

\section{Conclusion}
For portfolio optimization, this paper proposed a novel predictive probabilistic deep learning model that can represent the probability distribution of predictions. It was demonstrated that the proposed modification of ACGAN can be applied as a predictive model, although the existing ACGAN has been utilized as a generative model. Furthermore, PredACGAN has advantages in representing uncertainty in prediction with the proposed risk measurement using the entropy of the predictive distribution. Such an advantage of PredACGAN is distinct from the conventional predictive NNs, which cannot show the uncertainty of predictions. Additionally, a straightforward algorithm to use the risk measure for portfolio optimization was proposed. Measured predictive uncertainty was utilized as a threshold for filtering unreliable predictions in portfolio algorithms. Therefore, the constructed portfolios consist only of stocks that are expected to be highly profitable and reliable.

The proposed algorithm and PredACGAN were evaluated with an experimental scenario of investing in S\&P 500 stocks for ten years from 2011 to 2020, after training with 1990 to 2010 daily return data. As a result, PredACGAN showed superior performance compared to the portfolios without considering the risk measure. The proposed algorithm and PredACGAN also demonstrated stable performance, which is evaluated by maximum drawdown and Sharpe ratio. It is expected that the proposed framework with PredACGAN and the proposed algorithm can further expand the discussions in deep learning-based portfolio optimization.

\bibliographystyle{unsrt}  
\bibliography{ref}

\end{document}